\documentclass{article}%
\usepackage{amssymb}
\usepackage{amsmath}%
\usepackage{amsfonts}%
\usepackage{graphicx}

\begin{document}

\author{Vladimir K. Petrov\thanks{ E-mail address: vkpetrov@yandex.ru}}
\title{\textbf{Primordial function and ambiguity in its determination}}
\date{\textit{N. N. Bogolyubov Institute for Theoretical Physics}\\
\textit{\ National Academy of Science of Ukraine}\\
\textit{\ 252143 Kiev, Ukraine. 01.12.2003}}
\maketitle

\begin{abstract}
We study the possibility to reconstruct the primordial function for some
periodic function. The procedure includes an analytical continuation of a
discrete function for Fourier coefficients computation, that introduces an
ambiguity. To establish conditions under which no ambiguity appears, the
modification of Carlson theorem is suggested. In a case when no subsidiary
condition is imposed, ambiguity in the definition of primordial function
does not go beyond the functions which belong to space $\Omega^{\prime}$.

\end{abstract}

\section{Introduction}

In a previous work \cite{me} we have put in correspondence to any tempered
distribution $F\left(  \varphi\right)  $ some distribution
\begin{equation}
\widehat{\Sigma}F\left(  \varphi\right)  \equiv\sum_{n=-\infty}^{\infty
}F\left(  \varphi+2\pi n\right)  \equiv\widetilde{F}\left(  \varphi\right)
\label{per}%
\end{equation}
Recall that tempered distributions $F\left(  \varphi\right)  \in S^{\prime}$
are defined as functional
\begin{equation}
\left\langle F,\phi\right\rangle =\int_{-\infty}^{\infty}F\left(
\varphi\right)  \phi\left(  \varphi\right)  d\varphi
\end{equation}
on a space $S=\left\{  \phi\right\}  $ of fast decreasing test functions
$\phi$. Any $F\left(  \varphi\right)  \in S^{\prime}$ can be represented as
finite ($m<\infty$) derivative $F\left(  \varphi\right)  =\frac{\partial^{m}%
}{\partial\varphi^{m}}f\left(  \varphi\right)  $ of some continuous tempered
function ( $\left\vert f\left(  \varphi\right)  \right\vert <\left\vert
\varphi\right\vert ^{\sigma};\left\vert \varphi\right\vert \rightarrow\infty;$
$\sigma=const$) \cite{Schwartz,gel-shil,gel-shil 2,bremermann,vladimirov}.
Moreover, Fourier transform $%
\mathcal{F}%
\left[  F\left(  \varphi\right)  ,t\right]  $ of any $F\left(  \varphi\right)
\in S^{\prime}$ also belongs to space $S^{\prime}$ \cite{gel-shil
2,bremermann} and every periodic distribution is tempered \cite{donoghue}.

It is clear that $\widehat{\Sigma}$ is linear operator, i.e. if $\widehat
{\Sigma}f\left(  \varphi\right)  =\widetilde{f}\left(  \varphi\right)  $ and
$\widehat{\Sigma}g\left(  \varphi\right)  =\widetilde{g}\left(  \varphi
\right)  $, then $\widehat{\Sigma}\left[  \alpha f\left(  \varphi\right)
+\beta g\left(  \varphi\right)  \right]  =\alpha\widetilde{f}\left(
\varphi\right)  +\beta\widetilde{g}\left(  \varphi\right)  $.

Procedure $\left(  \ref{per}\right)  $, which we call the cyclization, is used
not infrequently (see e.g. \cite{Schwartz}), but as a rule cyclization is
applied to functions $F\left(  \varphi\right)  $ that decrease quite rapidly
with $\varphi\rightarrow\pm\infty$ and $\widetilde{F}\left(  \varphi\right)  $
appears periodic. In \cite{me} we considered the cases when this is not true
and $\widetilde{F}\left(  \varphi\right)  \neq\widetilde{F}\left(
\varphi+2\pi\right)  $. In this paper we handle only periodic $\widetilde
{F}\left(  \varphi\right)  $ and consider the possibility to reconstruct
function $F\left(  \varphi\right)  $ (called here 'primordial') from the given
$\widetilde{F}\left(  \varphi\right)  $, in other words, we consider procedure
$\widehat{\Sigma}^{-1}\widetilde{F}\left(  \varphi\right)  =F\left(
\varphi\right)  $ inverse to $\left(  \ref{per}\right)  $. Since there is a
family of functions $\omega\left(  \varphi\right)  $\ such that $\widehat
{\Sigma}\omega\left(  \varphi\right)  =0$, the inverse procedure is ambiguous
and we estimate the degree of possible ambiguity.

\section{Primordial functions}

We will perform the cyclization procedure for $F\left(  \varphi\right)  $
represented as the Fourier integral hence an interchange of integration and
summation order will be needed. Such interchange is justified in a case of
uniform convergence. It is clear that if $F_{t}\in S^{\prime}$, then the
Abel-Poisson regularization\ for Fourier integrals
\begin{subequations}
\begin{equation}
F\left(  \varphi\right)  =%
\mathcal{F}%
\left[  F_{t},\varphi\right]  =\int_{-\infty}^{\infty}F_{t}e^{it\varphi
}dt\rightarrow\int_{-\infty}^{\infty}F_{t}e^{it\varphi-\varepsilon\left\vert
t\right\vert }dt\label{f-i}%
\end{equation}
and Fourier series\footnote{Parameter $\varepsilon$ is a positive
parameter, which is regarded as arbitrary small but finite and, if the
opposite is not specified, it finally tends to zero.}
\end{subequations}
\begin{equation}
\widetilde{F}\left(  \varphi\right)  \equiv\sum_{n=-\infty}^{\infty}%
F_{n}e^{i\varphi n}\rightarrow\sum_{n=-\infty}^{\infty}F_{n}e^{i\varphi
n-\varepsilon\left\vert n\right\vert }\label{A-P}%
\end{equation}
provide a uniform convergence in $\left(  \ref{A-P}\right)  $ and $\left(
\ref{f-i}\right)  $. Abel-Poisson regularization\ also provides a standard
representation (see e.g. \cite{vladimirov}) for tempered distribution
\begin{subequations}
\begin{equation}
F\left(  \varphi\right)  =F_{+}\left(  \varphi+i\varepsilon\right)
-F_{-}\left(  \varphi-i\varepsilon\right)  ;\qquad\widetilde{F}\left(
\varphi\right)  =\widetilde{F}_{+}\left(  \varphi+i\varepsilon\right)
-\widetilde{F}_{-}\left(  \varphi-i\varepsilon\right)
\end{equation}
with$\qquad$%
\end{subequations}
\begin{equation}
F_{\pm}\left(  \varphi\right)  =\int_{0}^{\pm\infty}F_{t}e^{it\varphi
}dt;\qquad\widetilde{F}_{\pm}\left(  \varphi\right)  =\mp\frac{1}{2}F_{0}%
\pm\sum_{n=0}^{\pm\infty}F_{n}e^{i\varphi n}%
\end{equation}

Taking into account that
\begin{equation}
\sum_{m=-\infty}^{\infty}e^{2\pi tm}=\sum_{m=-\infty}^{\infty}\delta\left(
t-n\right) \label{del-P}%
\end{equation}
one can easily check that the application of procedure $\left(  \ref{per}%
\right)  $\ to regularized integral $\left(  \ref{f-i}\right)  $\ gives
\begin{equation}
\widetilde{F}\left(  \varphi\right)  =\sum_{n=-\infty}^{\infty}\int_{-\infty
}^{\infty}e^{i\varphi t}\delta\left(  t-n\right)  F_{t}dt\label{se}%
\end{equation}

The definition of product $\delta\left(  t-n\right)  F_{t}$\ for arbitrary
$F_{t}$\ goes beyond the distribution theory and will be partly considered
elsewhere. In this paper we consider only those $F_{t}$\ for which this
product exists in the distribution theory framework. Recall that Dirac's
$\delta$-function is defined as
\begin{equation}
\left\langle \delta\left(  t-s\right)  ,\phi\left(  t\right)  \right\rangle
=\int_{-\infty}^{\infty}\delta\left(  t-s\right)  \phi\left(  t\right)
dt=\phi\left(  s\right) \label{de-cont}%
\end{equation}
for any $\phi\left(  t\right)  $\ continuous in the small neighborhood of
point $t=s$.

After \cite{vladimirov} we call the distribution $F_{t}$\ continuous in the
area $\Delta$\ (i.e. $F_{t}\in C^{0}\left(  \Delta\right)  $) if there exists
some function $f_{t}$\ continuous in $\Delta$\ and%
\begin{equation}
\left\langle \left(  F_{t}-f_{t}\right)  ,\phi\left(  t\right)  \right\rangle
=\int\left(  F_{t}-f_{t}\right)  \phi\left(  t\right)  dt=0.\label{ft}%
\end{equation}
for any $\phi\left(  t\right)  \in S$\ with support located in $\Delta$. As a
$\Delta$\ we choose the set of small neighborhoods $\left(  n-\varepsilon
,n+\varepsilon\right)  =$\ $\Delta_{n,\varepsilon}$\ of all integer
$n$,\ where positive parameter $\varepsilon\ $is arbitrary small, but finite.
As it is shown in \cite{vladimirov}, expression $\left(  \ref{ft}\right)
$\ defines $f_{t}$\ uniquely almost everywhere. Further on $F_{t}$\ will be
interpreted as piecewise 'assembled' distribution which coincides with
original $F_{t}$\ for $t\notin$\ $\Delta_{n,\varepsilon}$\ and we put
$F_{t}=f_{t}$\ for $t\in$\ $\Delta_{n,\varepsilon}$. It is clear that in
within the distribution theory such 'assembled' distribution does not differ
from the original one.

A family of\ distributions $\left\{  F_{t}\right\}  =S_{\Delta}^{\prime}$
continuous in the area $\Delta$\ forms a linear space $S_{\Delta}^{\prime
}\subset S^{\prime}$. Indeed, $S_{\Delta}^{\prime}$\ includes zero element and
$\alpha_{1}F_{1,t}+\alpha_{2}F_{2,t}\in S_{\Delta}^{\prime}$\ for any
$F_{k,t}\in S_{\Delta}^{\prime}$\ and arbitrary complex numbers $\alpha_{k}$.
For any $F_{t}\in S_{\Delta}^{\prime}$\ integration in $\left(  \ref{se}%
\right)  $\ may be done and we get%
\begin{equation}
\widetilde{F}\left(  \varphi\right)  =\sum_{n=-\infty}^{\infty}\int_{-\infty
}^{\infty}e^{i\varphi t}\delta\left(  t-n\right)  f_{t}dt=\sum_{n=-\infty
}^{\infty}e^{i\varphi n}F_{n}%
\end{equation}
In this case $\widetilde{F}\left(  \varphi\right)  $ can be represented as
Fourier series $\left(  \ref{A-P}\right)  $ with coefficients given by the
standard expression%
\begin{equation}
F_{n}=\frac{1}{2\pi}\int_{-\pi}^{\pi}\widetilde{F}\left(  \varphi\right)
\exp\left\{  -i\varphi n\right\}  d\varphi\label{Fn}%
\end{equation}

The simplest way to compute primordial function $F\left(  \varphi\right)
$\ is the straighforward extension of function $F_{n}$\ in $\left(
\ref{Fn}\right)  $\ to the arbitrary value of parameter $n\rightarrow t$,
i.e.
\begin{equation}
\overline{F}_{t}=\frac{1}{2\pi}\int_{-\pi}^{\pi}\widetilde{F}\left(
\varphi\right)  \exp\left\{  -i\varphi t\right\}  d\varphi=\frac{1}{2\pi}%
\int_{-\infty}^{\infty}\overline{F}\left(  \varphi\right)  \exp\left\{
-i\varphi t\right\}  d\varphi\label{ext}%
\end{equation}
where%
\begin{equation}
\overline{F}\left(  \varphi\right)  \equiv\left\{
\begin{array}
[c]{cc}%
\widetilde{F}\left(  \varphi\right)  & -\pi\leq\varphi<\pi\\
0 & otherwise
\end{array}
\right. \label{trunc}%
\end{equation}

We wish to stress the difference between $\overline{F}_{t}$ and $F_{t}$.
Indeed,
\begin{eqnarray}
F_{t}=\frac{1}{2\pi}\int_{-\infty}^{\infty}e^{-i\varphi t}F\left(
\varphi\right)  d\varphi&=&\frac{1}{2\pi}\sum_{m=-\infty}^{\infty}\int_{\left(
2n-1\right)  \pi}^{\left(  2n+1\right)  \pi}e^{-i\varphi t}F\left(
\varphi\right)  d\varphi\nonumber \\&=&\frac{1}{2\pi}\int_{-\pi}^{\pi}e^{-i\varphi
t}\digamma\left(  \varphi,t\right)  d\varphi\label{f-t}%
\end{eqnarray}
where%
\begin{equation}
\digamma\left(  \varphi,t\right)  =\sum_{m=-\infty}^{\infty}e^{-i2n\pi
t}F\left(  \varphi+2n\pi\right)  \neq\widetilde{F}\left(  \varphi\right)
\end{equation}
This, in particular explains the difference in analytical properties of
$\overline{F}_{t}$ and $F_{t}$. Expression $\left(  \ref{ext}\right)  $ shows
that $\overline{F}_{t}$ is Fourier transform of distribution $\overline
{F}\left(  \varphi\right)  $ with compact support, hence $\overline{F}_{t}$ is
entire function. On the contrary, tempered distribution $F_{t}$ may be
singular. The only additional restriction is that $F_{t}$\ must be continuous
on a small neighborhoods of all integer $t=n$.

From $\left(  \ref{ext}\right)  $ and $\left(  \ref{per}\right)  $ we see that
$\widehat{\Sigma}\overline{F}\left(  \varphi\right)  =$ $\widetilde{F}\left(
\varphi\right)  $ so that $\overline{F}\left(  \varphi\right)  $ may be
regarded as one of the primordial function for $\widetilde{F}\left(
\varphi\right)  $. Another primordial function for $\widetilde{F}\left(
\varphi\right)  $ may be obtained, if instead $\left(  \ref{ext}\right)  $ one
takes
\begin{equation}
\overline{F}_{t}^{\left[  \sigma\right]  }=\frac{1}{2\pi}\int_{\sigma-\pi
}^{\sigma+\pi}\widetilde{F}\left(  \varphi\right)  e^{-i\varphi t}d\varphi
\end{equation}
where $\sigma$ is an arbitrary complex number. For all integer $t=n$ we get
$\overline{F}_{n}^{\left[  \sigma\right]  }=F_{n}$, but for arbitrary $t$ the
difference of extensions
\begin{equation}
\omega_{t}=\overline{F}_{t}^{\left[  \sigma\right]  }-\overline{F}_{t}%
\equiv\left(  e^{\pi it}-e^{-\pi it}\right)  \rho_{t}\label{om eg}%
\end{equation}
does not turn into zero, if
\begin{equation}
\rho_{t}=\frac{1}{2\pi}\int_{0}^{\sigma}\widetilde{F}\left(  \varphi
-\pi\right)  e^{-i\varphi t}d\varphi\neq0\label{ro-si}%
\end{equation}

Between $\overline{F}\left(  \varphi\right)  $ and $\widetilde{F}\left(
\varphi\right)  $ there exists a biunique correspondence. Despite the fact
that no Fourier series reproduces $\overline{F}\left(  \varphi\right)  $ for
$\left\vert \varphi\right\vert >\pi$, expression $\left(  \ref{Fn}\right)  $
does not allow to discriminate coefficients $F_{n}$ computed with
$\widetilde{F}\left(  \varphi\right)  $ from the ones computed with truncated
function $\overline{F}\left(  \varphi\right)  $. However, extension
$F_{n}\rightarrow\overline{F}_{t}$ with $\left(  \ref{ext}\right)  $ does not
obligatory coincide with $F_{t}$ given by $\left(  \ref{f-t}\right)  $.
Indeed, from $\left(  \ref{ext}\right)  $ one can easily get
\begin{equation}
\overline{F}_{t}=\frac{1}{2\pi}\int_{-\infty}^{\infty}\overline{F}\left(
\varphi\right)  e^{-i\varphi t}d\varphi=\sum_{n=-\infty}^{\infty}F_{n}%
\frac{\sin\pi\left(  n-t\right)  }{\pi\left(  n-t\right)  }\label{cont}%
\end{equation}
The difference between Fourier transforms of genuine and truncated
functions$\overline{F}$
\begin{equation}
\omega_{t}=\frac{1}{2\pi}\int_{-\infty}^{\infty}\left(  F\left(
\varphi\right)  -\overline{F}\left(  \varphi\right)  \right)  e^{-i\varphi
t}d\varphi\equiv F_{t}-\overline{F}_{t}=F_{t}-\sum_{n=-\infty}^{\infty}%
F_{n}\frac{\sin\pi\left(  n-t\right)  }{\pi\left(  n-t\right)  }\label{dif}%
\end{equation}
generally turns into zero only for integer $t$. So the main problem in
reconstruction of a primordial function is the estimation of admissible
ambiguity in analytical continuation of $F_{n}$.

\section{Uniqueness of straighforward extension}

If the class of functions is specified to which the required $F_{t}$ belongs ,
this will impose restrictions on the analytical extension of $F_{n}$ and may
reduce ambiguity in the reconstruction of $F_{t}$ from its values at integer
$t=n$. It is known, that $F_{n}$ may be \textit{uniquely }extended on the
whole complex plane \textit{, }if $F_{t}$ obeys the Carlson
theorem\footnote{There are Carlson theorem alternatives (see e.g.
\cite{bieber,hille}) that differ by a set of such conditions.} conditions.
According this theorem \cite{bieber}, if $F_{t}$ is an entire function and the
indicator (called also the growth-measurihng function) defined as%

\begin{equation}
h\left(  \alpha;F_{t}\right)  \equiv\overline{\lim}_{r\rightarrow\infty}%
\frac{1}{r}\ln\left\vert F_{c+r\exp\left(  i\alpha\right)  }\right\vert
\label{ind}%
\end{equation}
obeys the conditions
\begin{equation}
h\left(  \alpha;F_{t}\right)  <\lambda\label{h-b}%
\end{equation}
for some $\lambda=const$, and
\begin{equation}
h\left(  -\frac{\pi}{2};F_{t}\right)  +h\left(  \frac{\pi}{2};F_{t}\right)
<2\pi,\label{h-c}%
\end{equation}
then
\begin{equation}
F_{n}=0;\qquad n=0,1,2...\label{zeros}%
\end{equation}
leads to $F_{t}\equiv0$. In other words, any entire function $F_{t}$ is
defined uniquely by its values at integer $t=n=0,1,2...$, if $F_{t}$ obeys
conditions $\left(  \ref{h-b}\right)  $ and $\left(  \ref{h-c}\right)  $.

If one takes the expression $\left(  \ref{ext}\right)  $ for analytical
extension $F_{n}\rightarrow\overline{F}_{t}$, it will impose strong
restrictions on $\overline{F}_{t}$. Indeed, $\left(  \ref{ext}\right)  $ may
be interpreted as Fourier transform of distribution with a support located in
$-\pi\leq\varphi<\pi$, therefore \cite{gel-shil 2,bremermann} $\overline
{F}_{t}$ is an entire function and for $\left\vert t\right\vert \rightarrow
\infty$%
\begin{equation}
h\left(  \alpha;F_{t}\right)  \leq\pi\left\vert \sin\alpha\right\vert
\label{main}%
\end{equation}
so that condition $\left(  \ref{h-b}\right)  $ is satisfied. However, at the
same time $\left(  \ref{main}\right)  $ does not exclude
\begin{equation}
h\left(  -\frac{\pi}{2};F_{n}\right)  +h\left(  \frac{\pi}{2};F_{n}\right)
=2\pi,\label{W h-c}%
\end{equation}
hence condition $\left(  \ref{h-c}\right)  $ may be violated and Carlson
theorem may fail to work. Incidentally, entire function%
\begin{equation}
\sigma_{t}=\frac{1}{2\pi}\int_{-\pi}^{\pi}\left(  \sum_{n=-\infty}^{-1}%
\sigma_{n}\exp\left\{  i\varphi n\right\}  \right)  \exp\left\{  -i\varphi
t\right\}  d\varphi=\sum_{n=-\infty}^{-1}\sigma_{n}\frac{\sin\pi\left(
t-n\right)  }{\pi\left(  t-n\right)  }%
\end{equation}
obey $\left(  \ref{W h-c}\right)  $ and turns into zero for all $t=0,1,2...$,
but it does not lead to $\sigma_{t}\equiv0$.

One can easily check that if $\widetilde{F}\left(  \varphi\right)  $ turns
into zero at some finite interval $\left\vert \varphi-\varphi_{0}\right\vert
<\epsilon$; $0<\epsilon<2\pi$, the analytical extension $\left(
\ref{ext}\right)  $ acquires the form
\begin{equation}
\overline{F}_{t}=\frac{1}{2\pi}\int_{\epsilon}^{2\pi-\epsilon}\widetilde
{F}\left(  \varphi\right)  e^{-i\varphi t}d\varphi
\end{equation}
and condition $\left(  \ref{h-c}\right)  $ is satisfied. Such case, however,
is rather specific then routine.

Nonetheless, Carlson theorem may be easily modified for Fourier transforms of
distributions with support located on interval of the length $2\pi$. To be
more specific, we consider $\overline{F}\left(  \varphi\right)  $ defined in
$\left(  \ref{trunc}\right)  $. Expression $\left(  \ref{ext}\right)
$\ defines $F_{n}$\ uniquely. But if $F_{n}$\ is specified, series $\left(
\ref{A-P}\right)  $\ uniquely defines $\widetilde{F}\left(  \varphi\right)
$\ \cite{vladimirov} and consequently $\overline{F}\left(  \varphi\right)  $.
Thereby if for all integer $n$\ (positive and negative)%
\begin{equation}
F_{n}=0;\quad n=0,\pm1,\pm2..,\label{zeros_s}%
\end{equation}
then from $\left(  \ref{A-P}\right)  $ we get $\widetilde{F}\left(
\varphi\right)  =\overline{F}\left(  \varphi\right)  =0$, which with $\left(
\ref{ext}\right)  $ leads to $F_{t}\equiv0$.

\section{Linear space $\Omega$}

Here we estimate the maximum allowable ambiguity in computation of primordial
function when no subsidiary conditions are imposed. Any admissible extension
$F_{t}$ was to define Fourier coefficients $F_{n}$ uniquely. This, in
particular means that
\begin{equation}
F_{n}=\lim_{t\rightarrow n+0}F_{t}=\lim_{t\rightarrow n-0}F_{t}%
\end{equation}
i.e. $F_{t}$ was to be continuous in the area $\Delta$ defined above, so
$F_{t}$ is to belong to space $S_{\Delta}^{\prime}$. Evidently the same will
be true for the difference of any two extensions $F_{1,t}-F_{2,t}=\omega_{t}$.

A family of distributions $\left\{  \omega_{t}\right\}  \subset S_{\Delta
}^{\prime},$ such that $\omega_{n}=0$ for all integer $n$ forms linear space
$\Omega$.\ Indeed, $\Omega$ includes zero element and $\alpha_{1}\omega
_{1,t}+\alpha_{2}\omega_{2,t}\in\Omega$\ for any $\omega_{k,t}\in\Omega$\ and
arbitrary complex numbers $\alpha_{k}$.

It is clear that space $\Omega$\ may include only those entire functions of
exponential type which violate condition $\left(  \ref{h-c}\right)  $.
Moreover, the analytical extension of any $\omega_{n}\in\Omega$ with $\left(
\ref{ext}\right)  $ may give only trivial result $\omega_{t}\equiv0$, i.e. no
$\omega_{t}\in\Omega$ but $\omega_{t}\equiv0$ can be represented as $\left(
\ref{ext}\right)  $.

Fourier transform
\begin{equation}
\omega\left(  \varphi\right)  =\int_{-\infty}^{\infty}\exp\left\{  i\varphi
t\right\}  \omega_{t}dt\label{Fom}%
\end{equation}
relates space $\Omega=\left\{  \omega_{t}\right\}  $ to linear space
$\Omega^{\prime}=$ $\left\{  \omega\left(  \varphi\right)  \right\}  $. Note,
that if $\omega\left(  \varphi\right)  \in\Omega^{\prime}$ and $\varphi_{0}$
is a constant which does not spoil the convergence of integral in $\left(
\ref{Fom}\right)  $ (e.g. $\operatorname{Im}\varphi_{0}=0$), then
$\omega\left(  \varphi+\varphi_{0}\right)  \in\Omega^{\prime}$.

From $\left(  \ref{del-P}\right)  $ we obtain
\begin{equation}
\widehat{\Sigma}\omega\left(  \varphi\right)  \equiv\sum_{k=-\infty}^{\infty
}\omega\left(  \varphi+2\pi k\right)  =\sum_{m=-\infty}^{\infty}\int_{-\infty
}^{\infty}\exp\left\{  i\varphi t\right\}  \delta\left(  t-m\right)
\omega_{t}dt\label{nil 3}%
\end{equation}
Since $\omega_{t}$ is continuous at integer $t=n$, integration in $\left(
\ref{nil 3}\right)  $ may be done and, considering $\omega_{n}=0$, we get%
\begin{equation}
\widehat{\Sigma}\omega\left(  \varphi\right)  =\sum_{n=-\infty}^{\infty}%
\omega_{n}e^{i\varphi n}\equiv\widetilde{\omega}\left(  \varphi\right)
=0\label{nil}%
\end{equation}
i.e. procedure $\left(  \ref{per}\right)  $ turns any of $\omega\left(
\varphi\right)  \in\Omega^{\prime}$ into zero. $\ $In other words, space
$\Omega^{\prime}$ belongs to the kernel of the operator $\widehat{\Sigma}$.
Note, that for any distribution $\psi\left(  \varphi\right)  $ such that $%
\mathcal{F}%
^{-1}\left[  \psi\left(  \varphi\right)  ,t\right]  \in S_{\Delta}^{\prime}$,
the inverse proposition is true as well, i.e. $\widehat{\Sigma}\psi\left(
\varphi\right)  =0$ leads to $\psi\left(  \varphi\right)  \in\Omega^{\prime}$.

It should be noted that procedure $\left(  \ref{per}\right)  $ turns any
functions periodic with the period $2\pi\tau$ into zero, if $1/\tau$ is
noninteger number. Indeed, arbitrary function with the period $\tau$ may be
represented as%
\begin{equation}
\Phi\left(  \varphi,\tau\right)  =\sum_{n=-\infty}^{\infty}c_{n}\left(
1/\tau\right)  \exp\left\{  in\varphi/\tau\right\}
\end{equation}
so, taking into account $\left(  \ref{del-P}\right)  $ we get
\begin{equation}
\widehat{\Sigma}\exp\left\{  i\varphi/\tau\right\}  =\sum_{n=-\infty}^{\infty
}\exp\left\{  i\left(  \varphi-2\pi n\right)  /\tau\right\}  =\exp\left\{
i\varphi/\tau\right\}  \sum_{m=-\infty}^{\infty}\delta\left(  m-1/\tau\right)
=0
\end{equation}
Now we were to summarize over all admissible $\tau$, i.e. over all noninteger
$t=1/\tau$. It means that $c_{n}\left(  1/\tau\right)  \equiv c_{n}\left(
t\right)  \in\Omega$ and consequently
\begin{equation}
\sum_{n=-\infty}^{\infty}\int_{-\infty}^{\infty}c_{n}\left(  t\right)
\exp\left\{  in\varphi t\right\}  dt\equiv\sum_{n=-\infty}^{\infty}%
C_{n}\left(  \varphi\right)  \in\Omega^{\prime}%
\end{equation}
and we see that no additional ambiguity is introduced.

If $1/\tau$\ is integer, then $\widehat{\Sigma}\exp\left\{  i\varphi\right\}
=\exp\left\{  i\varphi\right\}  \coth\varepsilon/2\rightarrow\infty$\textrm{.
}Therefore, if some function $f\left(  \varphi\right)  $\ contains a term
periodic with the period $2\pi n$, then $\widehat{\Sigma}$\ convert $f\left(
\varphi\right)  \ $into an infinite constant. Since we reconstruct primordial
function $F\left(  \varphi\right)  $\ from periodic function $\widetilde
{F}\left(  \varphi\right)  $\ which does not contain infinite constants, such
terms are out of consideration.

For any $\omega\left(  \varphi\right)  \in\Omega^{\prime}$ one may define
\begin{equation}
\rho\left(  \varphi\right)  =\widetilde{\rho}\left(  \varphi\right)  +\rho
_{A}\left(  \varphi\right)  ;\quad\rho_{A}\left(  \varphi\right)  =\sum
_{k=0}^{\infty}\omega\left(  \varphi-\pi\left(  2k+1\right)  \right)
\label{rh-a}%
\end{equation}
where $\widetilde{\rho}\left(  \varphi\right)  $ is arbitrary periodic
distribution. With this definition one may write
\begin{equation}
\omega\left(  \varphi\right)  =\rho\left(  \varphi+\pi\right)  -\rho\left(
\varphi-\pi\right)  \;\label{am}%
\end{equation}
that leads to\footnote{In the particular case, considered above, $\omega_{t}$
is given by $\left(  \ref{om eg}\right)  $ and $\rho_{t}$ may be given by
$\left(  \ref{ro-si}\right)  $.}
\begin{equation}
\omega_{t}=\left(  e^{i\pi t}-e^{-i\pi t}\right)  \rho_{t}\label{om}%
\end{equation}

We claim that singularity of $\rho_{t}\equiv%
\mathcal{F}%
\left[  \rho\left(  \varphi\right)  ,t\right]  \in S^{\prime}$ at integer
$t=n$ is weaker than pole
\begin{equation}
\omega_{n}=2i\lim_{t\rightarrow n\pm0}\left(  \rho_{t}\sin\pi t\right)
=0\label{om2}%
\end{equation}
that guarantees $\omega_{t}\in\Omega$ and no additional restrictions on
$\rho_{t}$ are imposed.

Note, that application of the procedure $\left(  \ref{per}\right)  $\ to
$\rho\left(  \varphi\pm\pi\right)  $\ with arbitrary $\rho\left(
\varphi\right)  $\ does not obligatory convert these functions into periodic
ones \cite{me}. Therefore combination $\rho\left(  \varphi-\pi\right)
-\rho\left(  \varphi+\pi\right)  $\ does not obligatory turn into zero after
application of the procedure $\left(  \ref{per}\right)  $, but if condition
$\left(  \ref{om2}\right)  $\ is satisfied, it does.

The change of $e^{i\pi t}-e^{-i\pi t}$ in $\left(  \ref{om}\right)  $ for
arbitrary polynomial in $e^{i\pi t}$ and $e^{-i\pi t}$ which turns into zero
for all integer $t$ leads to nothing but redefinition of $\rho_{t}$. It is
clear also that e.g. $\omega_{m,t}=\left(  e^{i\pi t}-e^{-i\pi t}\right)
^{\frac{1}{m}}q_{t}\in\Omega$ for any finite $m$, if $\rho_{t}\equiv\left(
e^{i\pi t}-e^{-i\pi t}\right)  ^{\frac{1}{m}-1}q_{t}$ obeys $\left(
\ref{om}\right)  $.

Let tempered distribution $q_{t}$ be bounded for the integers $t=n$ and
$q_{t}\neq0$ at least for some integer $t$, i.e. $q_{t}\notin\Omega$. Since
\begin{equation}
\lim_{m\rightarrow\infty}\left\langle \phi\left(  t\right)  ,\left(  e^{i\pi
t}-e^{-i\pi t}\right)  ^{\frac{1}{m}}q_{t}\right\rangle =\left\langle
\phi\left(  t\right)  ,q_{t}\right\rangle
\end{equation}
for any $\phi\in S$, then, by definition sequence $\omega_{m,t}\in\Omega$.
However $\omega_{m,t}$ converges to $q_{t}\notin\Omega$ with $m\rightarrow
\infty$. In other words, $\Omega$ is an \textit{incomplete} space.

Let us define the 'window' operator $\widehat{\Omega}$ as
\begin{equation}
\widehat{\Omega}F_{t}\equiv\left\{
\begin{array}
[c]{cc}%
0 & t\in\Delta_{n,\varepsilon}\\
F_{t} & t\notin\Delta_{n,\varepsilon}%
\end{array}
\right.  ;\quad F_{t}\in S^{\prime}%
\end{equation}
which cuts out small neighborhoods $\Delta_{n,\varepsilon}$ of all integer
$t=n$. It is natural to define%
\begin{equation}
\widehat{\Omega}F\left(  \varphi\right)  \equiv\int_{-\infty}^{\infty}%
\widehat{\Omega}F_{t}\exp\left\{  i\varphi t\right\}  dt
\end{equation}
so for any $F\left(  \varphi\right)  \in S^{\prime}$ we get
\begin{equation}
\widehat{\Omega}F\left(  \varphi\right)  =\sum_{m=-\infty}^{\infty}%
\int_{m+\varepsilon}^{m+1-\varepsilon}F_{t}\exp\left\{  i\varphi t\right\}
dt\in\Omega^{\prime}%
\end{equation}
On the other hand, since
\begin{equation}
\left(  1-\widehat{\Omega}\right)  F_{t}=\left\{
\begin{array}
[c]{cc}%
F_{t}=f_{t} & t\in\Delta_{n,\varepsilon}\\
0 & t\notin\Delta_{n,\varepsilon}%
\end{array}
\right.  ;\quad F_{t}\in S_{\Delta}^{\prime}%
\end{equation}
with $f_{t}$ defined in $\left(  \ref{ft}\right)  $, one may write%
\begin{equation}
\left(  1-\widehat{\Omega}\right)  F\left(  \varphi\right)  =\int_{-\infty
}^{\infty}\left(  1-\widehat{\Omega}\right)  F_{t}\exp\left\{  i\varphi
t\right\}  dt=\sum_{m=-\infty}^{\infty}\int_{m+\varepsilon}^{m-\varepsilon
}F_{t}\exp\left\{  i\varphi t\right\}  dt\label{FF}%
\end{equation}
The continuity of $F_{t}$ on $\Delta_{n,\varepsilon}$ allows to apply the mean
value theorem and rewrite $\left(  \ref{FF}\right)  $ as
\begin{equation}
\left(  1-\widehat{\Omega}\right)  F\left(  \varphi\right)  =2\frac
{\sin\varphi\varepsilon}{\varphi}\sum_{m=-\infty}^{\infty}F_{m}e^{i\varphi
m}=\frac{2\sin\varphi\varepsilon}{\varphi}\widetilde{F}\left(  \varphi\right)
.
\end{equation}
It is clear that the distribution $\left(  1-\widehat{\Omega}\right)  F\left(
\varphi\right)  $ vanishes with $\varepsilon\rightarrow0$. Nonetheless,%
\begin{equation}
\left(  1-\widehat{\Omega}\right)  \sum_{n=-\infty}^{\infty}F\left(
\varphi+2\pi n\right)  =\widetilde{F}\left(  \varphi\right)  \sum_{n=-\infty
}^{\infty}\frac{2\sin\left(  \varphi+2\pi n\right)  \varepsilon}{\left(
\varphi+2\pi n\right)  }.
\end{equation}
hence, taking into account%
\begin{equation}
\sum_{n=-\infty}^{\infty}\frac{\exp\left\{  in\alpha\right\}  }{\varphi-2\pi
n}=-\frac{i}{2}\exp\left\{  i\varphi\frac{\alpha_{\operatorname{mod}2\pi}%
}{2\pi}\right\}  \left(  \operatorname*{signum}\left(  \alpha
_{\operatorname{mod}2\pi}\right)  +i\cot\left(  \frac{\varphi}{2}\right)
\right)  ,
\end{equation}
we find%
\begin{equation}
\sum_{n=-\infty}^{\infty}\frac{2\sin\left[  \left(  \varphi-2\pi n\right)
\varepsilon\right]  }{\left(  \varphi-2\pi n\right)  }=1
\end{equation}
and, consequently%
\begin{equation}
\left(  1-\widehat{\Omega}\right)  \widehat{\Sigma}F\left(  \varphi\right)
=\left(  1-\widehat{\Omega}\right)  \sum_{n=-\infty}^{\infty}F\left(
\varphi+2\pi n\right)  =\widetilde{F}\left(  \varphi\right)
\end{equation}

Since $\widehat{\Omega}^{2}=\widehat{\Omega}$, then both $\widehat{\Omega} $
and $1-\widehat{\Omega}$ may be considered as projectors. A family of tempered
distribution $\Xi=\left\{  \widehat{\Omega}\xi_{t}\right\}  $, with $\xi
_{t}\in S^{\prime}$, forms a linear subspace of $\Omega^{\prime} $. For
distributions $\lambda_{t}$ which are regular and continuous on all
$\Delta_{n,\varepsilon}$ one can define a family of functions $\Lambda
=\left\{  \left(  1-\widehat{\Omega}\right)  \lambda_{t}\right\}  $, which
forms a linear space as well. Cyclization procedure $\left(  \ref{per}\right)
$ converts any $\lambda\left(  \varphi\right)  =$ $%
\mathcal{F}%
\left[  \left(  1-\widehat{\Omega}\right)  \lambda_{t},\varphi\right]  $ into
a periodic function $\widehat{\Sigma}\lambda\left(  \varphi\right)
\equiv\widetilde{\lambda}\left(  \varphi\right)  =\widetilde{\lambda}\left(
\varphi+2\pi\right)  $.

Some linear transforms in spaces $\Omega$ and $\Omega^{\prime}$ are considered
in Appendix.

Now we turn to the $\omega_{t}$ specified in $\left(  \ref{dif}\right)  $ and
try to compute the corresponding function $\rho_{t}$. Taking into account%

\begin{equation}
\sum_{n=-\infty}^{\infty}\int_{-\pi}^{\pi}\exp\left\{  -i\alpha\left(
t-n\right)  \right\}  \frac{d\alpha}{2\pi}=\sum_{n=-\infty}^{\infty}\int
_{-\pi}^{\pi}\delta\left(  \frac{\alpha}{2\pi}-n\right)  \exp\left\{  -i\alpha
t\right\}  \frac{d\alpha}{2\pi}=1
\end{equation}
one may rewrite $\left(  \ref{dif}\right)  $ as%

\begin{equation}
\omega_{t}=\frac{\sin\pi t}{\pi}\sum_{n=-\infty}^{\infty}\frac{\overline
{F}_{t}-F_{n}}{t-n}\left(  -1\right)  ^{n}%
\end{equation}
that leads to
\begin{equation}
\rho_{t}=\frac{1}{2\pi i}\sum_{n=-\infty}^{\infty}\frac{\overline{F}_{t}%
-F_{n}}{t-n}\left(  -1\right)  ^{n}%
\end{equation}

\section{Conclusions}

In this paper we study a possibility to compute primordial function $F\left(
\varphi\right)  $ for some periodic function $\widetilde{F}\left(
\varphi\right)  $ represented by Fourier series $\left(  \ref{A-P}\right)
\ $and presumably being a result of application of procedure $\left(
\ref{per}\right)  $ to $F\left(  \varphi\right)  $.

The reconstruction of primordial function demands an analytical continuation
of Fourier coefficients, known only for integer $t=n$, to the whole real axis
$t$. Such extension will introduce an ambiguity, unless the class of functions
to which $F\left(  \varphi\right)  $ belongs is not essentially restricted.

As it is known, the extended function $F_{t}$ is unique only if it obeys the
conditions of Carlson theorem. In many cases such condition may appear too
restrictive. We suggested a modification of Carlson theorem conditions for the
case of extension given by $\left(  \ref{ext}\right)  $, which, as it seems to
us, is the simplest way to compute primordial function $F\left(
\varphi\right)  $.

In a case when no subsidiary condition is imposed, primordial function
$F\left(  \varphi\right)  $ may be reconstructed from $\widetilde{F}\left(
\varphi\right)  $ only up to arbitrary $\omega\left(  \varphi\right)
\in\Omega^{\prime}$.

\section{Appendix. Some linear transforms in spaces $\Omega$ and
$\Omega^{\prime}$}

The distribution theory defines product or convolution only for narrow class
of distributions. As it is stated in \cite{bremermann} any tempered infinitely
differentiable function $K_{t}$ is a multiplier in space $S$ so there exists a
product of \textit{such} $K_{t}$ with any tempered distribution. For some
subspace of $S^{\prime}$ restrictions on $K_{t}$ may appear to be weaker.

Let us assume that for some family of $\omega_{t}^{\left[  K\right]  }%
\in\Omega$ there exists a product with some distribution $K_{t}$. It is
natural to assume that $K_{t}\omega_{t}^{\left[  K\right]  }\in\Omega$. This
will be true, indeed, if $K_{t}$ is not 'too singular' at integers $t=n$, i.e.
$K_{t}$ and $\omega_{t}^{\left[  K\right]  }$ obey the condition
\begin{equation}
\lim_{t\rightarrow n+0}K_{t}\omega_{t}^{\left[  K\right]  }=\lim_{t\rightarrow
n-0}K_{t}\omega_{t}^{\left[  K\right]  }=0
\end{equation}
for all integer $n$.

Existence of product $K_{t}\omega_{t}^{\left[  K\right]  }$ means that
convolution of the Fourier transforms $%
\mathcal{F}%
\left[  \omega_{t}^{\left[  K\right]  },\varphi\right]  =\omega^{\left[
K\right]  }\left(  \varphi\right)  $ and $%
\mathcal{F}%
\left[  K_{t},\varphi\right]  =K\left(  \varphi\right)  $
\begin{equation}
\int_{-\infty}^{\infty}K\left(  \alpha-\varphi\right)  \omega^{\left[
K\right]  }\left(  \alpha\right)  d\alpha\equiv\widehat{K}\omega^{\left[
K\right]  }\left(  \varphi\right)
\end{equation}
exists.

The main convolution property allows to write
\begin{equation}
\widehat{\Sigma}\omega^{\left[  K\right]  }\left(  \varphi\right)
=\sum_{n=-\infty}^{\infty}\omega^{\left[  K\right]  }\left(  \varphi+2\pi
n\right)  =\int_{-\infty}^{\infty}K\left(  \alpha-\varphi\right)
\sum_{n=-\infty}^{\infty}\omega^{\left[  K\right]  }\left(  \alpha+2\pi
n\right)  d\alpha\label{S-O}%
\end{equation}
in other words%
\begin{equation}
\widehat{\Sigma}\widehat{K}\omega^{\left[  K\right]  }\left(  \varphi\right)
=\widehat{K}\widehat{\Sigma}\omega^{\left[  K\right]  }\left(  \varphi\right)
=0\quad\widehat{K}\omega^{\left[  K\right]  }\left(  \varphi\right)  \in
\Omega^{\prime}%
\end{equation}
that also follows from $K_{t}\omega_{t}^{\left[  K\right]  }\in\Omega$.

It is clear that a family of distributions $\Omega_{K}=\left\{  \omega
_{t}^{\left[  K\right]  }\right\}  $ forms a linear subspace of the space
$\Omega$. If in addition $\Omega_{K}$ includes all $K_{t}\omega_{t}^{\left[
K\right]  }$ then $\widehat{K}$ converts $\Omega_{K}^{\prime}=\left\{
\omega^{\left[  K\right]  }\left(  \varphi\right)  \right\}  $ into itself,
i.e. $\widehat{K}\Omega_{K}^{\prime}=\Omega_{K}^{\prime}$.

It is evident that in particular case%
\begin{equation}
K\left(  \varphi\right)  =H\left(  \varphi\right)  =\frac{1}{2\pi}\left(
\frac{1}{\varphi-i\varepsilon}+\frac{1}{\varphi+i\varepsilon}\right)
\end{equation}
the convolution%
\begin{equation}
\widehat{H}\omega^{\left[  H\right]  }\left(  \varphi\right)  =H\left[
\omega^{\left[  H\right]  },\varphi\right]  =\frac{1}{\pi}pv\int_{-\infty
}^{\infty}\frac{\omega^{\left[  H\right]  }\left(  \alpha\right)  }%
{\alpha-\varphi}d\alpha.\label{H-1}%
\end{equation}
may be interpreted as the Hilbert transform of $\omega^{\left[  H\right]
}\left(  \varphi\right)  $. Symbol $pv$ in $\left(  \ref{H-1}\right)  $ stands
for principal value of the integral.

From the fundamental property of the Hilbert transform we get
\begin{equation}
\widehat{H}\widehat{H}\omega^{\left[  H\right]  }\left(  \varphi\right)
=H\left[  H\left[  \omega^{\left[  H\right]  }\left(  \alpha\right)
,\beta\right]  \varphi\right]  =-\omega^{\left[  H\right]  }\left(
\varphi\right) \label{doble-Om}%
\end{equation}
So the operator $\widehat{H}$ converts subspace $\Omega_{H}^{\prime}%
\subset\Omega_{H}^{\prime}$ of functions $\omega^{\left[  H\right]  }$ which
admit the Hilbert transform into itself.

With the equation
\begin{equation}
\sum_{n=-\infty}^{\infty}\frac{1}{\varphi-2\pi n\pm i\varepsilon}\exp\left\{
-n\epsilon\right\}  =\pm\frac{i}{2}\mp\frac{i}{1-\exp\left\{  \pm i\left(
\varphi\pm i\varepsilon\right)  \right\}  }%
\end{equation}
one can directly check that $\widehat{\Sigma}H\left[  \omega^{\left[
H\right]  },\varphi\right]  =0$.

Recall that if $\omega^{\left[  H\right]  }\left(  \varphi\right)  $ is
'ordinary' function, then to belong to $\Omega_{H}^{\prime}$ it must be
piecewise continuously differentiable and there exists some constant
$\lambda>0$ such that for $\left\vert x\right\vert \rightarrow\infty$
\begin{equation}
\left\vert \omega^{\left[  H\right]  }\left(  \varphi\right)  \right\vert
=O(\left\vert \varphi\right\vert ^{-\lambda});\qquad\label{O-cond-H}%
\end{equation}

There are various approaches to Hilbert transform definition for
distributions. A brief review is given in \cite{brych-prud}. According to
\cite{gel-shil} the distributions $f\left(  \alpha\right)  $ which admit
Hilbert transform $H\left[  f\left(  \alpha\right)  ,\varphi\right]  $ may be
defined as the functional
\begin{align}
\left\langle H\left[  f\left(  \alpha\right)  ,\varphi\right]  ,\phi\left(
\varphi\right)  \right\rangle  & =\int_{-\infty}^{\infty}\left(  \frac{1}{\pi
}pv\int_{-\infty}^{\infty}\frac{\omega^{\left[  H\right]  }\left(
\alpha\right)  }{\alpha-\varphi}d\alpha\right)  \phi\left(  \varphi\right)
d\varphi\nonumber\\
& =\left\langle f\left(  \alpha\right)  ,H\left[  \phi\left(  \varphi\right)
,\alpha\right]  \right\rangle
\end{align}
on the test functions $\phi\left(  \varphi\right)  \in\Phi$, which are
infinitely differentiable and $\phi\left(  \varphi\right)  \leq O\left(
1/\left\vert \varphi\right\vert \right)  $ for $\left\vert \varphi\right\vert
\rightarrow\infty$. Taking into account $\left(  \ref{doble-Om}\right)  $ we
get $H\left[  \phi\right]  \in$ $\Phi$ for any $\phi\left(  \varphi\right)
\in$ $\Phi$, or $H\left[  \Phi\right]  =\Phi$. In particular%

\begin{equation}
\left\langle H\left[  \omega^{\left[  H\right]  }\right]  ,\phi\right\rangle
=\left\langle \omega^{\left[  H\right]  },\phi^{\prime}\right\rangle
,\qquad\phi^{\prime}\equiv H\left[  \phi\right]
\end{equation}
so $H\left[  \omega^{\left(  H\right)  }\right]  \in\Omega_{H}^{\prime}$. So
taking into account $\left(  \ref{doble-Om}\right)  $ we see that Hilbert
transform, indeed converts $\Omega_{H}^{\prime}$ into itself.

It should be noted that if condition $\left(  \ref{O-cond-H}\right)  $ is not
satisfied, Hilbert transform may be computed with some regularization. However
in this case an important property of the Hilbert transform
\begin{equation}
disc\left\{  H\left[  f\left(  \alpha\right)  ,\varphi\right]  \right\}
\equiv\lim_{\varepsilon\rightarrow0}\frac{H\left[  f\left(  \alpha\right)
,\varphi-i\varepsilon\right]  -H\left[  f\left(  \alpha\right)  ,\varphi
+i\varepsilon\right]  }{2i}=f\left(  \varphi\right) \label{disc}%
\end{equation}
may be broken. Indeed, Hilbert transform for $f\left(  \alpha\right)
=\exp\left\{  i\alpha\sigma\right\}  $%
\begin{equation}
H\left[  e^{i\alpha\sigma},\varphi\right]  =\frac{1}{\pi}pv\int_{-\infty
}^{\infty}\frac{e^{i\alpha\sigma}}{\sigma-\varphi}d\sigma=e^{i\alpha\varphi
}\frac{1}{\pi}pv\int_{-\infty}^{\infty}\frac{e^{i\alpha\sigma}}{\sigma}%
d\sigma=ie^{i\alpha\varphi}\operatorname*{signum}\left(  \alpha\right)
\end{equation}
is the entire function of $\varphi$ so $disc\left\{  H\left[  \exp\left\{
i\alpha\alpha\right\}  ,\varphi\right]  \right\}  \equiv0$.

For some functions that does not obey $\left(  \ref{O-cond-H}\right)  $ the
fundamental equation $\left(  \ref{doble-Om}\right)  $ may be broken too.
Indeed, Hilbert transform of $f=1$ is $H\left[  1\right]  =0$ (see e.g.
\cite{bateman}15.2$\left(  1\right)  $).

The reason is that we need the uniform convergence of the integral in $\left(
\ref{H-1}\right)  $ to change the order of integration in $\left(
\ref{doble-Om}\right)  $ or the order integration and proceeding to limit in
$\left(  \ref{disc}\right)  $. It is easy to prove that in the area
$\left\vert \varphi\right\vert <\Phi$ with arbitrary large but finite $\Phi$
integral $\left(  \ref{H-1}\right)  $ converge uniformly, if the condition
$\left(  \ref{O-cond-H}\right)  $ is satisfied.

\end{document}